\documentclass[12pt,epsf]{article}

\usepackage{scicite}

\input psfig.sty

\def\gsim{\mathrel{\rlap{\lower 4pt \hbox{\hskip 1pt $\sim$}}\raise 1pt
\hbox {$>$}}}
\def\lsim{\mathrel{\rlap{\lower 4pt \hbox{\hskip 1pt $\sim$}}\raise 1pt
\hbox {$<$}}}

\newenvironment{sciabstract}{%
\begin{quote} \bf}
{\end{quote}}

\newcounter{lastnote}

\title{
The first chemical enrichment in the universe and the formation
of hyper metal-poor stars } 
\author{
Nobuyuki Iwamoto,$^{1}$ 
Hideyuki Umeda,$^{2}$ 
Nozomu Tominaga,$^{2}$ \\
Ken'ichi Nomoto,$^{2\ast}$ 
Keiichi Maeda$^{3}$ \\
\\
\normalsize{$^{1}$Nuclear Data Center, Japan Atomic Energy Research Institute, Ibaraki 319-1195, Japan,}\\ 
\normalsize{$^{2}$Department of Astronomy, School of Science,  University of Tokyo, }\\
\normalsize{Tokyo 113-0033, Japan,}\\
\normalsize{$^{3}$Department of Earth Science and Astronomy,  College of Arts and Sciences, }\\
\normalsize{University of Tokyo, Tokyo 153-8902, Japan}\\
\\
\normalsize{$^\ast$To whom correspondence should be addressed; E-mail:
nomoto@astron.s.u-tokyo.ac.jp.} \\
\\
\normalsize{{\bf To be published in Science, and within the Science Express web site }}\\
\normalsize{{\bf (http://www.sciencexpress.org) on 2 June 2005}}
}

\date{}

\begin{document} 


\baselineskip24pt


\maketitle


\begin{sciabstract}
The recent discovery of a hyper metal-poor (HMP) star, whose metallicity
Fe/H is 
smaller than 1/100,000 of the solar ratio, together with one earlier
HMP star, has raised a challenging question if these HMP stars are the
actual first generation, low mass stars in the Universe.  We argue that
these HMP stars are the second generation stars being formed from
gases which were chemically enriched by the first generation
supernovae.  The key to this solution is the very unusual abundance
patterns of these HMP stars with important similarities and
differences. We can reproduce these abundance features with the
core-collapse ``faint'' supernova models which undergo extensive matter
mixing and fallback during the explosion.
\end{sciabstract}

 Identifying the first stars in the Universe, i.e., metal-free,
Population III (Pop III) stars which were born in a primordial
hydrogen-helium gas cloud is one of the important challenges of the
current astronomy~\cite{weiss00,abel02}.  Recently two hyper
metal-poor (HMP) stars, HE0107--5240~\cite{C02} and
HE1327--2326~\cite{F04}, were discovered, whose metallicity Fe/H is
smaller than 1/100,000 of the Sun (i.e., [Fe/H] $< -5$), being more
than a factor of 10 smaller than previously known extremely metal-poor
(EMP) stars.  (Here [A/B] $= \log_{10}(N_{\rm A}/N_{\rm B})-\log_{10}
(N_{\rm A}/N_{\rm B})_\odot$, where the subscript $\odot$ refers to
the solar value and $N_{\rm A}$ and $N_{\rm B}$ are the abundances of
elements A and B, respectively.)  This discovery was raised an
important question as to whether the observed low mass ($\sim$
0.8~$M_\odot$) HMP stars are actually Pop III stars, or whether these
HMP stars are the second generation stars being formed from gases
which were chemically enriched by a single first generation supernova
(SN)~\cite{UN03}.
This is related to the questions of how the initial mass function
depends on the metallicity~\cite{fsob03}.  Thus identifying the origin
of these HMP stars is indispensable to the understanding of the
earliest star formation and chemical enrichment history of the
Universe.

 The elemental abundance patterns of these HMP stars provide a key to
the answer to the above questions.  The abundance patterns of
HE1327--2326~\cite{F04} and HE0107--5240~\cite{C04,BESSELL04} are
quite unusual (Fig.~1). The striking similarity of [Fe/H] (=$-5.4$ and 
$-5.2$ for HE1327--2326 and HE0107--5240, respectively) and [C/Fe]
($\sim +4$) suggests that similar chemical enrichment mechanisms
operated in forming these HMP stars.  However, the N/C and (Na, Mg,
Al)/Fe ratios are more than a factor of 10 larger in HE1327--2326.  In
order for the theoretical models to be viable, these similarities and
differences should be explained self-consistently.

 Here we report our findings that the above similarities and
variations of the HMP stars can be well reproduced in unified manner
by nucleosynthesis in the core-collapse ``faint'' supernovae (SNe)
which undergo mixing-and-fallback~\cite{UN03}.  We thus argue that the
HMP stars are the second generation low mass stars, whose formation
was induced by the first generation (Pop III) SN with efficient
cooling of carbon-enriched gases.

 The similarity of [Fe/H] and [C/Fe] suggests that the progenitor's
masses of Pop III SNe were similar for these HMP stars.  We therefore
choose the Pop III 25 $M_\odot$ models and calculate their evolution
and explosion.  The abundance distribution after explosive
nucleosynthesis is shown in Figure~2 for the kinetic energy $E$
of the ejecta $E_{51} \equiv E/10^{51}~{\rm erg} = 0.74$.  The
abundance distribution for $E_{51} = 0.71$ is similar.  In the
``faint'' SN model, most part of materials that underwent explosive
nucleosynthesis are decelerated by the influence of the gravitational
pull~\cite{WW95} and will eventually fall back onto the central
compact object (Fig.~3).  Such ``fallback'' was not calculated in
ref.~\cite{UN03}, but is found to take place in the present modeling
if $E_{51} < 0.71$.  (For the 50~$M_\odot$ star, the fallback is found
to occur for $E_{51} < 2$ because of deeper gravitational potential.)
We obtain a relation between $E$ and the mass cut $M_{\rm cut}$ (the
mass of the materials which finally collapse to form a compact object),
i.e., smaller $E_{51}$ leads to a larger amount of fallback (larger
$M_{\rm cut}$).  The explosion energies of $E_{51} = 0.74$ and $0.71$
lead to the mass cut $M_{\rm cut} = 5.8 M_\odot$ and $6.3 M_\odot$,
respectively, and we use the former and the latter models to explain
the abundance patterns of HE1327--2326 and HE0107--5240, respectively.

 During the explosion, we assume that the SN ejecta undergoes mixing,
i.e., materials are first uniformly mixed in the mixing-region
extending from $M_r = 1.9 M_\odot$ to the mass cut at $M_r = M_{\rm
cut}$ (where $M_r$ is the mass coordinate and stands for the mass
interior to the radius $r$) as indicated in Figure~2 (also see
legend), and only a tiny fraction, $f$, of the mixed material is
ejected from the mixing-region together with all materials at $M_r >
M_{\rm cut}$; most materials interior to the mass cut fall back onto
the central compact object.  Such a mixing-fallback mechanism (which
might mimic a jet-like explosion) is required to extract Fe-peak and
other heavy elements from the deep fallback region into the
ejecta~\cite{UN03,UN05}.

 Figure~1 shows the calculated abundance ratios in the SN ejecta
models for suitable choice of $f$ (see legend of Fig.~2) which are
respectively compared with the observed abundances of the two HMP
stars.  To reproduce [C/Fe] $\sim$ +4 and other abundance ratios of
HMP stars in Figure~1, the ejected mass of Fe is only 1.0 $\times
10^{-5}M_\odot$ for HE1327--2326 and 1.4 $\times 10^{-5}M_\odot$ for
HE0107--5240 (see legend of Fig. 2 for other abundances).  These
SNe are much fainter in the radioactive tail than the
typical SNe and form massive black holes of $\sim 6 M_\odot$.

 The question is what causes the large difference in the amount of
Na-Mg-Al between the SNe that produced HE0107--5240 and HE1327--2326.
Because very little Na-Mg-Al is ejected from the mixed
fallback materials (i.e., $f \sim 10^{-4}$) compared with the
materials exterior to the mass cut, the ejected amount of Na-Mg-Al is
very sensitive to the location of the mass cut.  As indicated in
Figure~2, $M_{\rm cut}$ is smaller (i.e., the fallback mass is
smaller) in the model for HE1327--2326 ($M_{\rm cut} = 5.8 M_\odot$)
than HE0107--5240 ($M_{\rm cut} = 6.3 M_\odot$), so that a larger
amount of Na-Mg-Al is ejected from the SN for HE1327--2326.  Since
$M_{\rm cut}$ is sensitively determined by the explosion energy, the
(Na-Mg-Al)/Fe ratios among the HMP stars are predicted to show
significant variations and can be used to constrain $E_{51}$.  
Note also that the explosion energies of these SN models with fallback
are not necessarily very small (i.e., $E_{51} \sim 0.7$).
Further these explosion energies are consistent with those observed in
the actual "faint" SNe~\cite{Tura98}.

 Here we should note that our previous models~\cite{UN03} tend to
underproduce Na compared with the abundances of HE0107--5240.  This
problem has been improved in our new presupernova models.  Na and Al
are mainly produced by C shell-burning, and their production is very
sensitive to the treatment of overshooting in the convective C burning
shell as well as the $^{12}$C abundance left after core He
burning~\cite{CL02}.  By including overshooting with the overshooting
length less than one-fifth of a pressure scale height for whole
presupernova evolution, our new supernova models contain large enough
abundances of Na and Al as seen in Figure~2.  Such an overshooting
length has been estimated from the comparison with the HR diagrams of
many young stellar clusters.  After the mixing-and-fallback, the
resultant abundance patterns with Na and Al are in reasonable
agreement with HE1327--2326 and HE0107--5240 (Fig.~1).  The
enhancement of Na and Al attributable to overshooting in the
progenitor evolution may better explain the small odd-even effect in
the elemental abundance patterns observed in EMP
stars~\cite{cayrel04}.

 The next question is why HE1327--2326 has a much larger N/C ratio
than HE0107--5240.  In our models, a significant amount of N is
produced by the mixing between the He convective shell and the H-rich
envelope during the presupernova evolution~\cite{UNN00}, where C
created by the triple-$\alpha$ reaction is burnt into N through the CNO
cycle.  For the HE1327--2326 model, we assume about 30 times larger
diffusion coefficients (i.e., faster mixing) for the H and He
convective shells to overcome an inhibiting effect of the mean
molecular weight gradient (and also entropy gradient) between H and He
layers.  Thus, larger amounts of protons are carried into the He
convective shell.  Then [C/N] $\sim 0$ is realized as observed in
HE1327--2326.  Such an enhancement of mixing efficiency has been
suggested to take place in the present-day massive stars known as fast
rotators, which show various N and He enrichments due to different
rotation velocities~\cite{Heger00}.

 If no large enhancement of N occurred in the SN ejecta, the following
scenario can explain high abundance of N (and also Na and Al) at the
surface.  If HE1327--2326 is in a binary system and its companion star
had experienced the asymptotic giant branch (AGB) phase, only the
odd-elements such as N, Na and Al can be efficiently enriched.  The
observed C, Mg and heavier elements should predominantly come from a
faint SN as modeled above.  The small accreted mass (e.g., $\sim
10^{-4} - 10^{-3} M_\odot$) mixed with shallow surface convective
layer in HE1327--2326 is enough to account for the observed abundance
pattern.  The smallness of the accreted mass requires that the
observed star belongs to a wide binary system and accretion takes
place through mass loss from an AGB star.  In contrast to the
AGB-scenario without the pre-enrichment from the faint SN (see below),
this model can realize [C/N] $\sim 0$ if the proper amount of N is
transfered.

 For HE0107--5240, an alternative scenario has been proposed, assuming
that the HMP stars are actually Pop III stars.  Here we point out that
such a scenario has difficulty in explaining the differences between
HE0107--5240 and HE1327--2326.  This scenario assumes that the HMP
star is in a binary system and an AGB companion star has polluted the
surface abundance of the HMP star~\cite{S04}.  Even if Pop III AGB
stars suffer any surface pollution at the early phase of their
evolution~\cite{S04,CHI01,SLL02}, recurrent mixing-process after He
shell flashes (third dredge-up) carries C-enriched materials (with no
enrichment of N) from a deep He-rich layer to the surface .  The
surface C abundance of the low-mass companion progressively increases,
but no N enhancement can be seen (i.e., [C/N] $> 0$).  On the
contrary, if a donor AGB star experiences hot bottom burning,
dredged-up C is processed into N at the base of the convective
envelope and thus [C/N] $= -2 \sim -1$.  Therefore [C/N] $\approx 0$
is difficult to be reproduced by Pop III AGB stars, although the C/N
ratio might be consistent with the observed value during a short
period of the evolution.

 What about stars with 130-300 $M_\odot$~\cite{UN02,HW02} ?
Pair-instability SNe (PISNe) from this mass range have been widely
considered to be the first source of chemical enrichment in the
universe~\cite{HW02}.  However, PISNe provide abundance patterns that
are incompatible with the observations of the HMP stars.  Since PISNe
undergo complete disruption and eject a large amount of
Fe~\cite{UN02,HW02}, the ejecta have [C/Fe] that is too low ($< 0$) to
be compatible with the two HMP stars and large [Fe/H] (say $>-4$) is
predicted.

 What other elements are important to distinguish the different models?
Oxygen is certainly important.  For HE0107--5240, its large [C/O] ratio
rules out the simple mass-cut models (without mixing-fallback) in the
multiple SN model~\cite{BESSELL04,UN05}.  For our faint SN models,
[C/O] is sensitive to $M_{\rm cut}$ and thus $E$.  

 Neutron-capture elements are important for constraining scenarios
involving an AGB star.  For HE1327--2326, the observed lower limit
of [Sr/Ba] $>-0.4$ is inconsistent with the s-process enhanced
stars~\cite{AOKI01,LUC03} and theoretical predictions of low metallicity
AGB s-process~\cite{GM00}, but is remarkably consistent with the values
seen in the r-process enhanced stars~\cite{HILL02,SNEDEN03}.  This may
favor SN origins because the r-process signature observed in EMP stars
is thought to come from SNe, but one should recall that the s-process
in AGB stars is still uncertain and such a Sr/Ba ratio might also be
reproduced this way~\cite{SIESS04}.

 Our models offer several predictions for future observations of HMP
stars. (1) The metallicity Fe/H of an HMP star is determined by the
mass ratios between the ejected Fe $M_{\rm Fe}$ and mixed interstellar
H $M_{\rm Hmix}$, and small $M_{\rm Fe}$ (i.e., small $f$) is
responsible for the small [Fe/H].  Our spherical explosion models
predict a continuous distribution of [Fe/H] in metal-poor stars.
Thus, if the gap at [Fe/H] $\sim -5$ to $-4$ is real, jet-induced
mixing might be responsible for constraining the distribution of the
$f$-value.  (2) Assuming that C/H needs to be higher than a certain
value in order to form low-mass HMP stars, C/Fe would tend to be
larger for smaller Fe/H.  (3) The (Na-Mg-Al)/Fe ratios in HMP stars
would show a continuous distribution because their variations are the
result of variation of $E$.  (4) If the large N/Fe is attributable to
rotation and if rotation can contribute to enhance $E$, N/Fe would
show a positive correlation with (Na-Mg-Al)/Fe.

\clearpage
\begin{figure}
\psfig{file=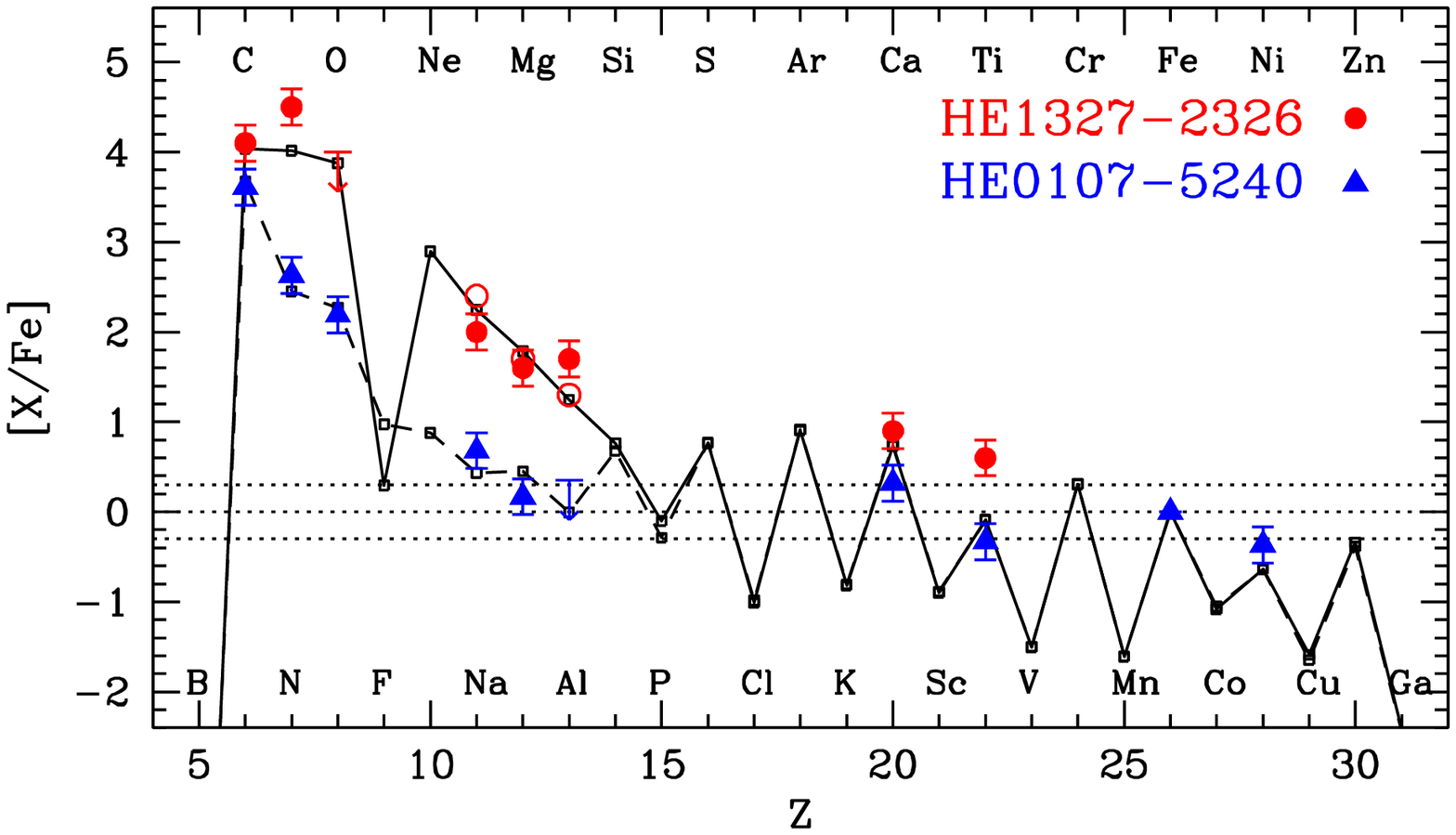,width=0.8\textwidth}
\caption{ Comparison of elemental abundance ratios observed in
HE1327--2326 [filled circles~\cite{F04}] and HE0107--5240 [filled
triangles~\cite{C04,BESSELL04}] with those of our supernova models (small
open squares connected by the solid line for HE1327--2326 and by the
dashed line for HE0107--5240) as a function of atomic number~$Z$ [here
the new solar abundances are used~\cite{ma04}].  For Na and Al, the
importance of accurate non-local thermodynamic equilibrium (LTE)
corrections are demonstrated from the comparison with the LTE values
indicated by the open circles.  The ejected yields are those from Pop
III 25~$M_\odot$ SN models whose parameters are given in the legend of
Figure 2.}
\end{figure}

\clearpage
\begin{figure}
\psfig{file=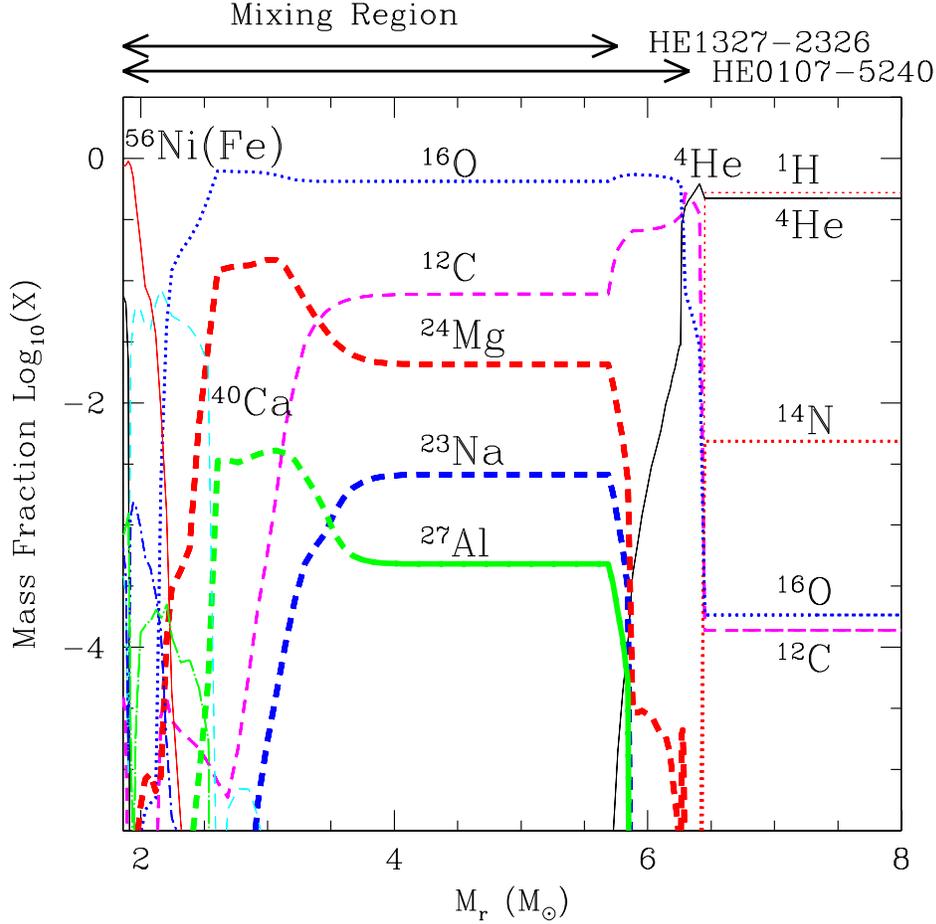,width=0.8\textwidth}
\caption{ Internal abundance distribution for nuclei (by mass
fraction) in the Pop III 25$M_\odot$ SN model for the
explosion energy of $E_{51}=0.74$ (i.e., for HE1327--2326).  The
distribution is similar for $E_{51}=0.71$ (HE0107--5240).  The mixing
is assumed to take place in the region of $M_r=1.9-5.8 M_\odot$ for
HE1327--2326, and $M_r=1.9-6.3 M_\odot$ for HE0107--5240.  The mass
fraction of the ejected materials with respect to the mixed fallback
materials is $f = 8.7 \times 10^{-5}$ for HE1327--2326, and $f = 1.2
\times 10^{-4}$ for HE0107--5240.  As a result, the ejecta contains
$1.0 \times 10^{-5} M_\odot$ $^{56}$Ni and 0.20 $M_\odot$ $^{12}$C for
HE1327-2326, and $1.4 \times 10^{-5}M_\odot$ $^{56}$Ni and 0.12
$M_\odot$ $^{12}$C for HE0107--5240.}
\end{figure}

\clearpage
\begin{figure}
\psfig{file=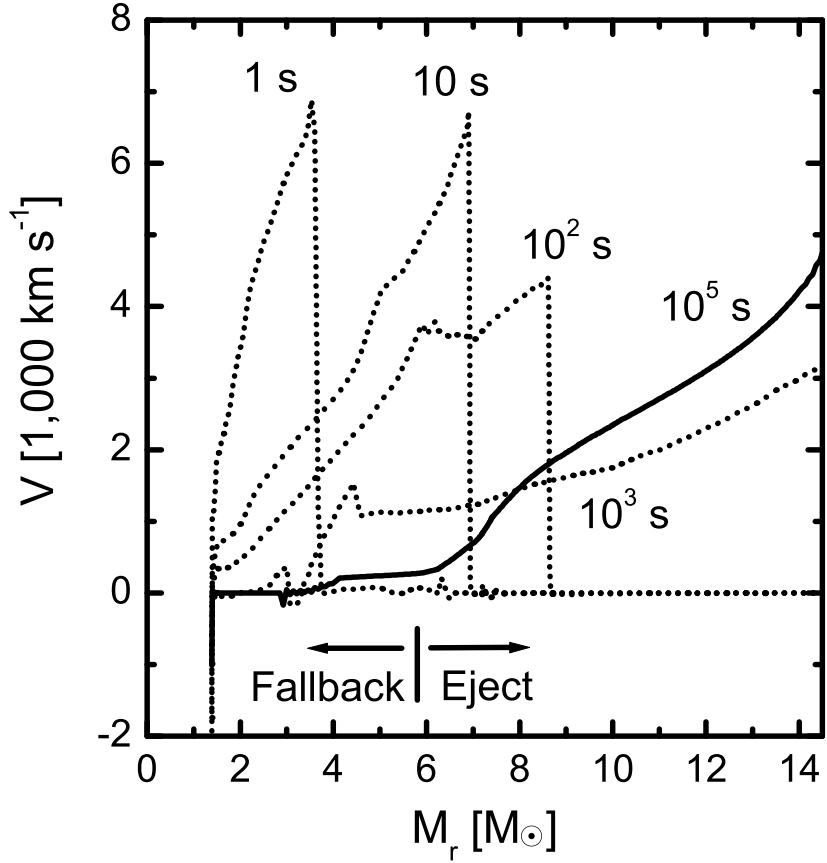,width=0.8\textwidth}
\caption{ Propagation of the shock wave and the fallback
of the model for 
HE1327--2326.  The progenitor is the 25 $M_\odot$ star.  As the shock
propagates through the H envelope and breaks out of the
surface, the materials in the inner region continue to be decelerated
and will eventually fallback onto the central remnant.  The mass cut
(that divides the materials fallen onto the central remnant and
ejected outward) is determined by comparing the velocity and the
escape velocity at $10^5$ seconds after the explosion.}
\end{figure}

\end{document}